\documentclass[superscriptaddress]{revtex4}
\usepackage{epsfig}
\usepackage{amsbsy}
\usepackage{amsmath}
\usepackage{amsfonts}
\usepackage{graphicx}

\begin{document}

\newcommand{\nn}{\nonumber}
\newcommand{\dg}{^\dagger}
\newcommand{\ip}[1]{\left\langle{#1}\right\rangle}
\newcommand{\bra}[1]{\langle{#1}|}
\newcommand{\ket}[1]{|{#1}\rangle}
\newcommand{\braket}[2]{\langle{#1}|{#2}\rangle}
\newcommand{\st}[1]{\langle {#1} \rangle}
\newcommand{\etal}{{\it et al}}
\newcommand{\half}{{\small \frac 12}}

\title{Quantum teleportation and entanglement swapping for spin systems}
\author{Dominic W Berry}
\affiliation{Department of Physics and
	Centre for Advanced Computing --- Algorithms and Cryptography,	\\
	Macquarie University,
	Sydney, New South Wales 2109, Australia}
\author{Barry C Sanders}
\affiliation{Department of Physics and
	Centre for Advanced Computing --- Algorithms and Cryptography,	\\
	Macquarie University,
	Sydney, New South Wales 2109, Australia}
\affiliation{Quantum Entanglement Project, ICORP, JST, Edward L Ginzton
 Laboratory, \\ Stanford University, CA 94305-4085, USA}
\date{\today}

\begin{abstract}
We analyse quantum teleportation (QT) and entanglement swapping (ES) for spin
systems. If the permitted operations are restricted to the Ising interaction,
plus local rotations and spin measurements, high-fidelity teleportation is
achievable for quantum states that are close to the maximally weighted spin
state. ES is achieved, and is maximized for a combination of entangled states
and Bell measurements that is different from the QT case. If more general local
unitary transformations are considered, then it is possible to achieve perfect
teleportation and ES.
\end{abstract}
\maketitle

\renewcommand{\thesection}{\arabic{section}}
\section{Introduction}
\label{sec:introduction}
Quantum teleportation (QT) enables disembodied transport of the state of a
system to a distant system through (i)~a shared entanglement resource, (ii)~a
classical communication channel between the sender and receiver \cite{Bennett}
and (iii)~an experimentally established isomorphism between the Hilbert spaces
of the sender and receiver \cite{Enk01}. QT is significant in several areas,
including transmission of quantum states in noisy
environments \cite{Bennett}, sharing states in distributed quantum
networks \cite{Cir99} and implementation of quantum computation using resources
prepared offline \cite{Got99,KLM}. Teleportation was initially proposed for
discrete-variable systems, where the state to be teleported has finite-$N$
levels \cite{Bennett}, and a continuous-variable (CV) version \cite{continuous}
has been adapted for squeezed light experiments \cite{CVQT}. Our interest here
is in QT of a quantum state in an arbitrary but finite $N$-dimensional Hilbert
space~${\cal H}_N$, realized physically as a spin system, thereby generalizing
the recent spin QT proposal by Kuzmich and Polzik (KP) which is only valid in
the infinite-$N$ limit \cite{Kuzmich}.

Entanglement swapping (ES) is closely related to QT. Whereas QT enables the
state of a system (e.g.\ a particle or collection of particles) to be
teleported to an independent physical system via classical communication
channels and a shared entanglement resource, the purpose of ES is to instill
entanglement between systems that hitherto shared no entanglement. An
entanglement resource is required for ES to occur; indeed the nomenclature
`entanglement swapping' describes the transfer of entanglement from
{\it a priori} entangled systems to {\it a priori} separable systems.

A connection between ES and QT can be seen as follows. Consider QT of the state
of one particle, which is initially entangled with a second particle, but the
state of the second particle does not undergo QT. In perfect QT the state of the
first particle is faithfully transferred to a third particle that was initially
independent of the first two particles. Thus, subsequent to the QT, the second
and third particles are entangled, perfectly replacing the {\it a priori}
entanglement of the first and second particles. The entanglement resource
inherent in QT devices enables this ES to occur; thus, equivalence between
optimal entanglement resources for QT and ES might be expected, but we show here
that the optimal entanglement resources differ between QT and ES for finite-$N$
spin systems.

QT of states in~${\cal H}_N$ can, in principle, be accomplished using the
entanglement resource 
\begin{equation}
\label{resce}
\ket{\Phi} = N^{-1/2} \sum_{m=1}^N \ket m \otimes \ket m \in
{\cal H}_N^{\otimes 2} \equiv {\cal H}_N\otimes{\cal H}_N,
\end{equation}
by employing the Bell state projective-valued measurement (PVM)
$\vert p,q \rangle\langle p,q \vert$ \cite{Bennett}, where the Bell states are
\begin{equation}
\label{bell}
\ket{p,q} = N^{-1/2} \sum_{m=1}^N {\rm e}^{{\rm i} 2\pi mp/N} \ket m \otimes
\ket{m+q \!\!\!\! \mod N}.
\end{equation}
These Bell states are mutually orthogonal and are all maximally entangled (see
appendix \ref{append} for further discussion of these states). This entanglement
resource and Bell measurement can also be used to perform ES for multilevel
systems by teleportation of entanglement. This PVM does not have an obvious
physical realization, however, except for the well-studied case where $N=2$.

One way of performing Bell measurements is via a two-mode unitary transformation
followed by measurements on the individual modes. It is possible, in principle,
to obtain the required unitary transformation from any available interaction and
local operations \cite{interact}. Here we consider the case where each mode is a
spin system, and the available interaction is the Ising interaction. If it were
possible to perform arbitrary local operations, then it would be possible to
perform perfect teleportation; this is considered further in section
\ref{sec:perfect}. Unfortunately it is not physically realistic to consider
arbitrary local operations for spin systems (except for the trivial spin-1/2
case). Therefore we consider the case where the local operations are restricted
to rotations. In addition we only consider local spin measurements, which are
the most physically realistic for spin systems.

A model of how to perform teleportation with these restrictions is provided by
the teleportation protocol of Kuzmich and Polzik \cite{Kuzmich}. They consider
the case that the entanglement resource is two entangled beams of light, with
reduced fluctuations in sums and differences of the components of the Stokes
vectors. The Stokes vector is equivalent to spin, as it obeys the same
commutation relations, though it does not physically correspond to spin. One of
these beams is passed through a sample of atoms, where the off-resonant
atom-photon interaction provides an interaction that is equivalent to the Ising
interaction. Measurement of a component of the Stokes vector of the light and of
the spin of the sample of atoms then provides the Bell measurement. Then an
appropriate rotation of the Stokes vector of the other beam of light gives
teleportation of the initial spin state of the atomic sample.

Here we apply this teleportation scheme, which is for the limit of infinite
spin, to the case of finite spin. We identify the optimal two-mode spin-squeezed
states for use as an entanglement resource in section \ref{sec:optimal} and
determine the fidelity of teleportation using this entanglement resource and the
Bell measurements of \cite{Kuzmich} in section \ref{sec:teleportation}. We
determine the level of ES achieved by this teleportation scheme in section
\ref{sec:entanglement} and determine a modified interaction that provides
significantly improved ES. In section \ref{sec:perfect} we identify a Bell state
PVM, adapted from that of section \ref{sec:entanglement}, such that perfect QT
and ES are obtained. Conclusions are presented in section \ref{sec:conclusions}.

\section{Two-mode spin-squeezed states}
\label{sec:optimal}

One of the three key criteria of QT listed in section \ref{sec:introduction} is
a shared entanglement resource. In order to establish suitable entangled states,
we determine two-mode spin-squeezed states via an optimization scheme and show
that these states are indeed entangled. We adapt the KP two-mode spin-squeezed
state (subject to a minor transformation of variables equivalent to a
permutation of indices for the spin operators) for which the two-mode standard
deviations satisfy
\begin{equation}
\label{uncs}
\Delta J_z^{(+)} \sim 0,~\,~~\Delta J_y^{(-)} \sim 0 \, .
\end{equation}
Here $J_k^{(\pm)} = J_k^{(1)} \pm J_k^{(2)}$, where $J_k^{(1)}$ and $J_k^{(2)}$
are the spin operators for the two modes 1 and 2, for $k \in \{x,y,z\}$. Each of
the two systems is represented in the same $j$-irrep; consequently a basis set
for two-mode squeezed states is
$\{\vert jm\rangle_z\otimes\vert jn\rangle_z\}\in{\cal H}_{2j+1}^{\otimes 2}$,
where $\vert jm\rangle_z\otimes\vert jn\rangle_z$ is a simultaneous eigenstate
of $J_z^{(1)}$ and $J_z^{(2)}$, with corresponding eigenvalues $m$ and $n$
respectively. An arbitrary state can be expressed as
\begin{equation} \label{expand}
\ket \Phi = \sum_{mn} \Phi_{mn} \ket{jm}_z \otimes \ket{jn}_z.
\end{equation}

In order for the KP teleportation scheme to be effective, it is also necessary
that $\st{J_x^{(+)}} \lesssim 2j$. This requirement of proximity to the
highest-weight state of $J_x^{(+)}$ requires a compromise with the squeezing
condition~(\ref{uncs}) because decreasing the uncertainties in (\ref{uncs})
decreases $\st{J_x^{(+)}}$ as well, by analogy with the single-mode
spin-squeezed state counterpart \cite{Kit,Sor01}. Restricting to states near
the highest-weight state of $J_x^{(+)}$ is equivalent to restricting to states
near the $J_x^{(k)}$ highest-weight states for the individual modes $k=1$ and
2. This corresponds to working in the Heisenberg--Weyl (HW) limit of SU(2)
dynamics, i.e.\ the dynamics is close to that of a harmonic oscillator
\cite{Rowe}.

In order to obtain two-mode spin-squeezed states that give high fidelity for
the KP teleportation scheme, we need to balance the conflicting criteria of
equation (\ref{uncs}) and $\st{J_x^{(+)}} \lesssim 2j$. This can be done by
minimizing the quantity
\begin{equation}
\label{chimin}
\chi(\mu) = V_z^{(+)}+V_y^{(-)}-\mu \st{J_x^{(+)}},
\end{equation}
where $V_z^{(+)}$ and $V_y^{(-)}$ are the variances of $J_z^{(+)}$ and
$J_y^{(-)}$ respectively. This optimization gives the minimal variances
$V_z^{(+)}$ and $V_y^{(-)}$ for the maximum value of $\st{J_x^{(+)}}$, and the
value of $\mu$ weights the relative importance of minimizing the variances as
compared to maximizing $\st{J_x^{(+)}}$. The optimum value of $\mu$ will be
found numerically.

It is only possible to minimize $\chi$ numerically, as it is fourth order in the
state coefficients $\Phi_{mn}$ of equation (\ref{expand}). We have performed
this numerical minimization from random initial states \cite{randsta} for spins
up to $j=5$, and found that both $\st{J_z^{(+)}}$ and $\st{J_y^{(-)}}$ are equal
to zero for the optimal states. We conjecture that
$\st{J_z^{(+)}}=\st{J_y^{(-)}}=0$ for the optimal states for arbitrary spin.
This implies that $V_z^{(+)}=\st{(J_z^{(+)})^2}$ and
$V_y^{(-)}=\st{(J_y^{(-)})^2}$ for the optimal states. 

When we make this replacement in equation (\ref{chimin}), the optimization
problem reduces to the minimization of the expectation value of an operator.
Using the method of undetermined multipliers, this optimization can be performed
by solving the eigenvalue equation
\begin{equation}
\label{eigeq}
\Big[ (J_z^{(+)})^2+(J_y^{(-)})^2 -\mu J_x^{(+)} \Big] \ket{\nu;\mu} = \nu
\ket{\nu;\mu}.
\end{equation}
This optimization corresponds to determining the state that minimizes $\Delta
J_z^{(+)}$ and $\Delta J_y^{(-)}$, with the auxiliary constraint of a fixed
value of $\st{J_x^{(+)}}$.

The value of $\nu$ can take, in principle, values between $-2j\mu$ and
$8j^2+2j\mu$. It does not actually reach these bounds, however, because it is
not possible to have spectral extrema for all three terms on the left-hand side
simultaneously.  The actual bounds are only determined numerically. The
eigenstate corresponding to the minimum eigenvalue $\nu$ (i.e.\ closest to
$-2j\mu$) will be the optimal state. In our calculations we have found that this
eigenstate is unique, although it is not obvious that this should be the case.

We have performed numerical minimizations of $\chi$, using initial states found
by solving the eigenvalue equation (\ref{eigeq}), for spins up to $j=20$. These
states also minimize $\chi$ and satisfy $\st{J_z^{(+)}}=\st{J_y^{(-)}}=0$. This
vindicates our conjecture that $\st{J_z^{(+)}}=\st{J_y^{(-)}}=0$ for the optimal
states.

This optimization procedure is similar to the optimization for the single-mode
case considered by S\o rensen and M\o lmer \cite{Sor01}. They consider the
problem of minimizing the variance in $J_x$ for maximal $\st{J_z}$. In their
case there is the additional complication that for half-odd-integer spin,
$\st{J_x}$ is not necessarily zero for the optimal state, whereas we have
conjectured and verified numerically that the means are zero in the two-mode
case. This complication arises in the single-mode case because $J_x$ has
eigenvalues at $\pm 1/2$ but not at zero. It is therefore not possible for the
variance of $J_x$ to be less than $1/4$ if the mean is still zero. The
minimum-uncertainty states where the variance is less than $1/4$ must be
asymmetric with
a mean near $\pm 1/2$. On the other hand, for integer spin, $J_x$ has an
eigenvalue at zero, and the minimum uncertainty state is symmetric and has zero
mean, so the square is the same as the variance. This is equivalent to the
result that we find here, because the operators that we wish to minimize,
$J_z^{(+)}$ and $J_y^{(-)}$, have integer eigenvalues from $-2j$ to $+2j$,
including 0 (regardless of whether the spin $j$ is an integer or half an odd
integer).

\begin{figure}
\centering
\includegraphics[width=0.45\textwidth]{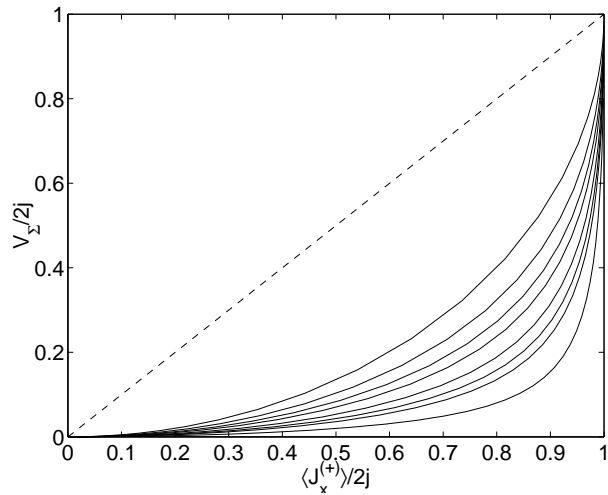}
\caption{Variances $V_\Sigma$ for optimal two-mode spin-squeezed states as a
function of $\st{J_x^{(+)}}$. In order of decreasing heights, the solid curves
correspond to $j=1/2$, 1, $3/2$, 2, 3, 4, 5, and 10. The dashed line corresponds
to $\chi(1)=0$.}
\label{spnsqz}
\end{figure}

In figure \ref{spnsqz} the results of optimization by the undetermined
multiplier technique are depicted as a graph of the sum of the variances for the
optimal state $V_\Sigma \equiv V_z^{(+)}+V_y^{(-)}$ versus the mean
$\st{J_x^{(+)}}$. These results are normalized by a factor of $2j$, in order to
better compare the results for different $j$. As can be seen, in order to obtain
smaller $V_\Sigma$, we require smaller values of $\st{J_x^{(+)}}$. It is
possible to obtain a state that satisfies (\ref{uncs}) perfectly, i.e.\
$V_\Sigma = 0$, at the expense of also having $\st{J_x^{(+)}}=0$. This state,
\begin{equation}
\label{maxent}
\ket{\mu=0;\nu=0} = \frac 1{\sqrt{2j+1}}\sum_{m=-j}^j \ket{jm}_y \otimes
\ket{jm}_y \, ,
\end{equation}
is maximally entangled and is equivalent to the entanglement resource of
equation (\ref{resce}) expressed for a two-mode spin system.

Thus far, the analysis has been focused on two-mode spin squeezing, whereas the
QT requirement is in fact entanglement. The two-mode system is entangled if
$\chi(1)<0$. To see this, recall that the commutation relation $[J_y,J_z]=i J_x$
implies that $V_y V_z \ge |\st{J_x}|^2/4$. In turn this implies that
\begin{equation}
\label{singineq}
V_z+V_y \ge V_z + \frac{|\st{J_x}|^2}{4V_z} \ge |\st{J_x}|.
\end{equation}
For unentangled pure states we have $V_z^{(+)}=V_z^{(1)}+V_z^{(2)}$ and
$V_y^{(-)}=V_y^{(1)}+V_y^{(2)}$. Note that this does not necessarily hold in the
case of a mixed state, as we may have classical correlations. Thus, for
unentangled pure states we may add the inequalities (\ref{singineq}) for the two
modes 1 and 2, giving
\begin{equation}
V_z^{(+)}+V_y^{(-)} \ge |\st{J_x^{(1)}}|+|\st{J_x^{(2)}}| \ge \st{J_x^{(+)}}.
\end{equation}
Hence $\chi(1) \ge 0$ for unentangled pure states, and $\chi(1)<0$ implies
entanglement. It is also possible to generalize this result to mixed states
using the method of Duan \etal\ \cite{Duan}.

The limit below which the states must be entangled, $\chi(1)=0$, is depicted in
figure \ref{spnsqz} as a dashed line. Except for the end points at
$\st{J_x^{(+)}}/2j=0$ or 1, all the results are below this line, demonstrating
entanglement. The converse does not hold, however: entanglement does not
necessarily imply $\chi(1)<0$. For example, the maximally entangled state
(\ref{maxent}) does not satisfy this inequality.

\section{Quantum teleportation}
\label{sec:teleportation}

In order to perform QT, one share, or mode, of the entanglement resource will be
mixed by Alice with the unknown state~$\vert\psi\rangle\in{\cal H}_N$ to be
teleported in such a way that Alice should learn nothing about the state to be
teleported yet obtains classical measurement results $(a,b)$ (via a Bell-type
measurement) to share with Bob. Bob receives the results of Alice's measurement
via a classical channel. He then performs a unitary transformation based on the
result $(a,b)$ on the second mode of the two-mode entangled spin-squeezed state.
Bob's output state, designated $\ket{\zeta_{a,b}}$, should ideally be a replica
of the unknown input state $\ket \psi$.

The two modes of the entanglement resource are designated $1$ and $2$, and mode
$3$ is the state to be teleported. Alice's measurement will take place jointly
on modes $2$ and $3$, and Bob's approximate replica of the original state will
be in output mode~$1$. The Hilbert space for the three modes is
${\cal H}_{2j+1}^{\otimes 3}$. KP proposed the measurement corresponding to the
non-linear transformation \cite{Kuzmich}
\begin{equation}
\label{interact}
U=\exp({\rm i}J_y^{(2)}J_z^{(3)}/j),
\end{equation}
followed by a joint measurement of $J_z^{(2)}$ and $J_y^{(3)}$. Note that we
have omitted $\otimes$ in the above expression; we will omit the tensor product
symbol from this point on for the sake of brevity.

As mentioned in the previous section, we restrict to entangled states near the
maximally weighted $J_x$-eigenstate for modes 1 and 2 (the entanglement
resource). Similarly we apply the same restriction to the state to be
teleported. By `near' we mean that the state has significant support on the
states $\ket{jm}_x$ only for $m \sim j$. For this restriction, we can determine
the approximate effect of this unitary transformation using a contraction to
HW(2). To see this, consider the Holstein--Primakoff representation
\cite{Holstein}
\begin{align}
J_0 = j-\hat n, ~~~~
J_+ = \sqrt{2j-\hat n} \, a, ~~~~
J_- = a \dg \sqrt{2j-\hat n},
\end{align}
with
\begin{equation}
J_0 \equiv J_x, ~~~~ J_\pm \equiv J_y \pm {\rm i} J_z .
\end{equation}
The limit that we consider here is equivalent to the SU(2)$\to$HW(2)
contraction, as considered in \cite{Rowe}, with $\bar m\to j$. The operators
have the asymptotic forms
\begin{equation}
\label{asymptot}
J_0 \to j-a\dg a, ~~~~ J_+ \to \sqrt{2j} \, a, ~~~~ J_- \to \sqrt{2j} \, a \dg .
\end{equation}
In this limit we also find that
\begin{align}
J_y &\to \sqrt{\frac j2} (a+a\dg) = \sqrt j \, x, \\
J_z &\to \frac 1{\rm i} \sqrt{\frac j2} (a-a\dg) = \sqrt j \, p.
\end{align}
Therefore, the unitary transformation (\ref{interact}) becomes
\begin{equation}
U\to U_{\rm HW} = \exp({\rm i}x^{(2)}p^{(3)}),
\end{equation}
which generates the displacements
\begin{equation}
\tilde p^{(2)} = p^{(2)}+p^{(3)}, ~~~~ \tilde x^{(3)} = x^{(3)}-x^{(2)} \, ,
\end{equation}
where the tildes indicate the transformed variables. Returning to the notation
of spin operators, we have the approximate transformation
\begin{equation}
\tilde J_z^{(2)} \approx J_z^{(2)}+J_z^{(3)},~~~~\tilde J_y^{(3)} \approx
J_y^{(3)}-J_y^{(2)} \, .
\end{equation}

Let us denote the measured values of $\tilde J_z^{(2)}$ and $\tilde J_y^{(3)}$
by $a$ and $b$ respectively. The two-mode squeezed state satisfies
equation (\ref{uncs}); hence,
\begin{equation}
{J_z'}^{(1)} \approx J_z^{(3)}-a,~~~~{J_y'}^{(1)} \approx J_y^{(3)}-b,
\end{equation}
where the prime indicates the operator subsequent to the measurement yielding
the result $(a,b)$. Following this measurement, Bob applies the unitary
transformation
\begin{equation}
\label{simple}
V(a,b) \equiv \exp[{\rm i}(a J_y^{(1)}-b J_z^{(1)})/j] .
\end{equation}

Using the HW(2) contraction, $V(a,b)$ becomes
\begin{equation}
V(a,b) \to V_{\rm HW}(a,b) = \exp[{\rm i}(a x^{(1)}-b p^{(1)})/\sqrt j],
\end{equation}
which gives the transformations
\begin{align}
V\dg_{\rm HW}(a,b) p^{(1)}V_{\rm HW}(a,b) = p^{(1)}+a/\sqrt j \, , \nn \\
V\dg_{\rm HW}(a,b) x^{(1)}V_{\rm HW}(a,b) = x^{(1)}+b/\sqrt j \, .
\end{align}
The approximate transformations of the spin operators are therefore
\begin{align}
V\dg(a,b) {J_z'}^{(1)} V(a,b) \approx {J_z'}^{(1)}+a , \nn \\
V\dg(a,b) {J_y'}^{(1)} V(a,b) \approx {J_y'}^{(1)}+b .
\end{align}
Therefore, the rotation $V(a,b)$ approximately negates the translations of $a$
and $b$. This is equivalent to the rotations considered by KP.

\begin{figure}
\centering
\includegraphics[width=0.45\textwidth]{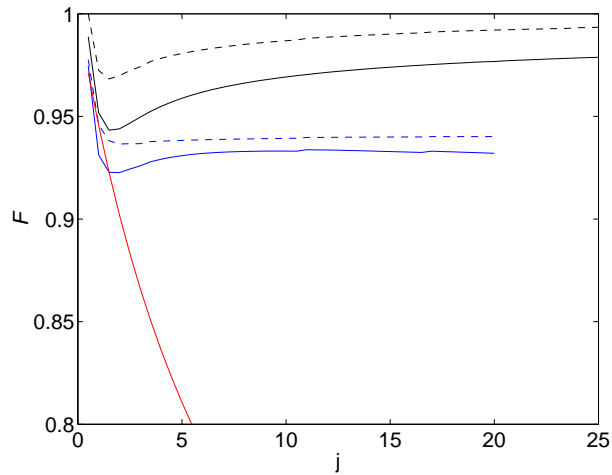}
\caption{Fidelity~$\cal F$ for teleportation of $\ket{jj}_x$ states (black
curves) and average fidelity ${\cal F}_{\rm av}$ over an ensemble of states near
$\ket{jj}_x$ (blue curves) as a function of $j$. The case where the final
transformation is as in equation (\ref{simple}) is shown as the continuous
curves, and that using (\ref{explicit}) is shown as the dashed curves. The red
curve is the approximate maximum fidelity without entanglement.}
\label{fidel}
\end{figure}

Fidelity~(\ref{eq:fidelity}) is used to characterize the QT of individual
states. As discussed in appendix~\ref{apfid}, we consider fidelity averaged over
the measurement results $(a,b)$, with weighting according to the probability of
obtaining these measurement results. This teleportation protocol is only
accurate for input states near the highest-weight eigenstates $\ket{jj}_x$, and
it should be most accurate for $\ket{jj}_x$. The fidelity for teleportation of
these states is plotted as a function of $j$ in figure \ref{fidel}. The
entanglement resource is the two-mode spin-squeezed state derived from equation
(\ref{eigeq}), with $\mu$ optimized to maximize $\cal F$ for teleportation.

The rotation $V(a,b)$ only approximately negates the terms $a$ and $b$, and
there are many other combinations of rotations that do this. For example, a
rotation can be made about the $z$-axis followed by a rotation about the
$y$-axis (${\rm e}^{{\rm i}aJ_y^{(1)}/j}{\rm e}^{-{\rm i}bJ_z^{(1)}/j}$), or
vice versa, or even a sequence of such rotations
(e.g.\ $[{\rm e}^{{\rm i}aJ_y^{(1)}/4j}{\rm e}^{-{\rm i}bJ_z^{(1)}/4j}]^4$).
In general, these rotations are equivalent to a rotation about an axis in the
$y$--$z$ plane (although with slightly different coefficients), plus an
additional rotation about the $x$-axis. As there should not be any additional
rotation about the $x$-axis, the most accurate teleportation should be for $V$
in the above form: a single rotation about an axis in the $y$--$z$ plane.

Nevertheless, this argument does not eliminate the possibility that more
accurate teleportation may be achieved by using the rotation $V$ with
coefficients slightly different from $a$ and $b$, particularly for the larger
values of these two variables. In general, the fidelity should be close to the
maximum possible if the orientation of the expectation value of the spin vector
for the teleported state is in the same direction as that for the initial state.
That is,
\begin{equation}
\st{V\dg {J_k'}^{(1)} V} = \gamma \st{J_k^{(3)}} \, ,
\end{equation}
for some proportionality constant $\gamma$ and $k\in \{x,y,z\}$. This
preservation of orientation can be achieved by using a suitable rotation about
an axis in the $y$--$z$ plane. Unfortunately this rotation will be dependent on
the input state, which is in general unknown in QT experiments. To avoid this
problem, we determined the rotations for a $\ket{jj}_x$ input state, and applied
these rotations to all input states. That is, the rotations $V$ used were those
that would give $\ip{J_y}=\ip{J_z}=0$ for the output state if the input state
were $\ket{jj}_x$. Explicitly this rotation is
\begin{equation}
\label{explicit}
V(j \xi \st{{J_y'}^{(1)}},j \xi \st{{J_z'}^{(1)}}) ,
\end{equation}
where
\begin{equation}
\xi =  \frac {\arccos \big(\st{{J_x'}^{(1)}}/|{\bf J}^{(1)}|\big)}
{\sqrt{\st{{J_y'}^{(1)}}^2+\st{{J_z'}^{(1)}}^2}}.
\end{equation}
for
\begin{equation}
|{\bf J}^{(1)}|^2 = \st{{J_x'}^{(1)}}^2+\st{{J_y'}^{(1)}}^2+\st{{J_z'}^{(1)}}^2.
\end{equation}
Here the expectation values implicitly depend on the measurement results
$(a,b)$, so these rotations also depend on $a$ and $b$.

For this scheme of final rotations, the fidelity is increased as the
entanglement resource is less entangled (for a $\ket{jj}_x$ input state). This
higher fidelity does not correspond to better teleportation, as the fidelity is
worse for states other than $\ket{jj}_x$. Therefore, rather than separately
determining optimum values of $\mu$ for the rotations (\ref{explicit}), the same
values as were determined previously for the $V(a,b)$ rotation were used. The
rotations (\ref{explicit}) give fidelities significantly higher than those for
the simple $V(a,b)$ rotation for $\ket{jj}_x$ states.

In general the quality of a teleportation scheme cannot be judged from the
fidelity of teleportation for just one state; we must consider the fidelity as a
function of the input state. A good way of quantifying the quality of the
teleportation scheme is to determine the average fidelity over an ensemble of
states, as in equation (\ref{fidav}). Analogously to \cite{Braun} for the
continuous case, we will take a weighted average over coherent spin states, with
weighting function
\begin{equation}
\label{prob}
W(\ket{\theta,\phi}) \propto {\rm e}^{-\theta^2/\sigma} ,
\end{equation}
where $\sigma$ is approximately the variance for the distribution. The state
$\ket{\theta,\phi}$ is a coherent spin state rotated an angle $\theta$ away from
the $x$-axis, that is,
\begin{equation}
\label{defcoh}
\ket{\theta,\phi} = {\rm e}^{{\rm i}\phi J_x} {\rm e}^{{\rm i}\theta J_y}
\ket{jj}_x \, .
\end{equation}
These coherent states can alternatively be expressed as
\begin{equation}
\ket{\theta,\phi} = {\rm e}^{{\rm i}\theta (J_y \cos \phi - J_z \sin \phi)}
\ket{jj}_x \, .
\end{equation}
In the limits (\ref{asymptot}), this becomes
\begin{equation}
\ket{\theta,\phi} \to {\rm e}^{{\rm i}\theta \sqrt{\frac j2}({\rm e}^{{\rm i}
\phi} a + {\rm e}^{-{\rm i}\phi} a\dg)} \ket 0 = D\left({\rm i}\theta
{\rm e}^{-{\rm i}\phi}\sqrt{\tfrac j2}\right) \ket 0 ,
\end{equation}
where $\ket 0$ is the harmonic oscillator vacuum state and $D(\cdots)$ is the
displacement operator. Thus we see that in this limit the coherent spin states
are equivalent to coherent states with coherent amplitude ${\rm i}\theta
{\rm e}^{-{\rm i}\phi}\sqrt{j/2}$.

Therefore, the probability distribution (\ref{prob}) is analogous to that used
in \cite{Braun}, with $\lambda \equiv 2/\sigma j$. It was found in
\cite{Braun} that the maximum fidelity without entanglement is
$(1+\lambda)/(2+\lambda)$. This means that the corresponding limit here is
approximately
\begin{equation}
\label{limit}
{\cal F}_{\rm max}(\sigma) \approx \tfrac 12 \frac{\sigma j+2}{\sigma j+1}.
\end{equation}
This will be a good approximation for small $\sigma$. Note that this approaches
$1/2$ in the limit of large $j$, which is the same as the limit for CV
teleportation \cite{Braun}.

The results as determined via numerical integrals, as well as this limit, are
depicted in figure \ref{fidel}. These results are for the example of
$\sigma=(20^\circ)^2$. The average fidelities are still high, well above $0.9$,
but do not tend to 1 for large spin. Instead they tend towards asymptotic values
slightly below 1. Note also that the more sophisticated scheme of final
rotations (\ref{explicit}) again gives higher fidelity than the simple case
$V(a,b)$. The fidelity in both of these cases is well above the limit
(\ref{limit}) for spin above about 2. Similar results are obtained for other
values of $\sigma$, with the average fidelity decreasing as $\sigma$ is
increased.

\begin{figure}
\centering
\includegraphics[width=0.45\textwidth]{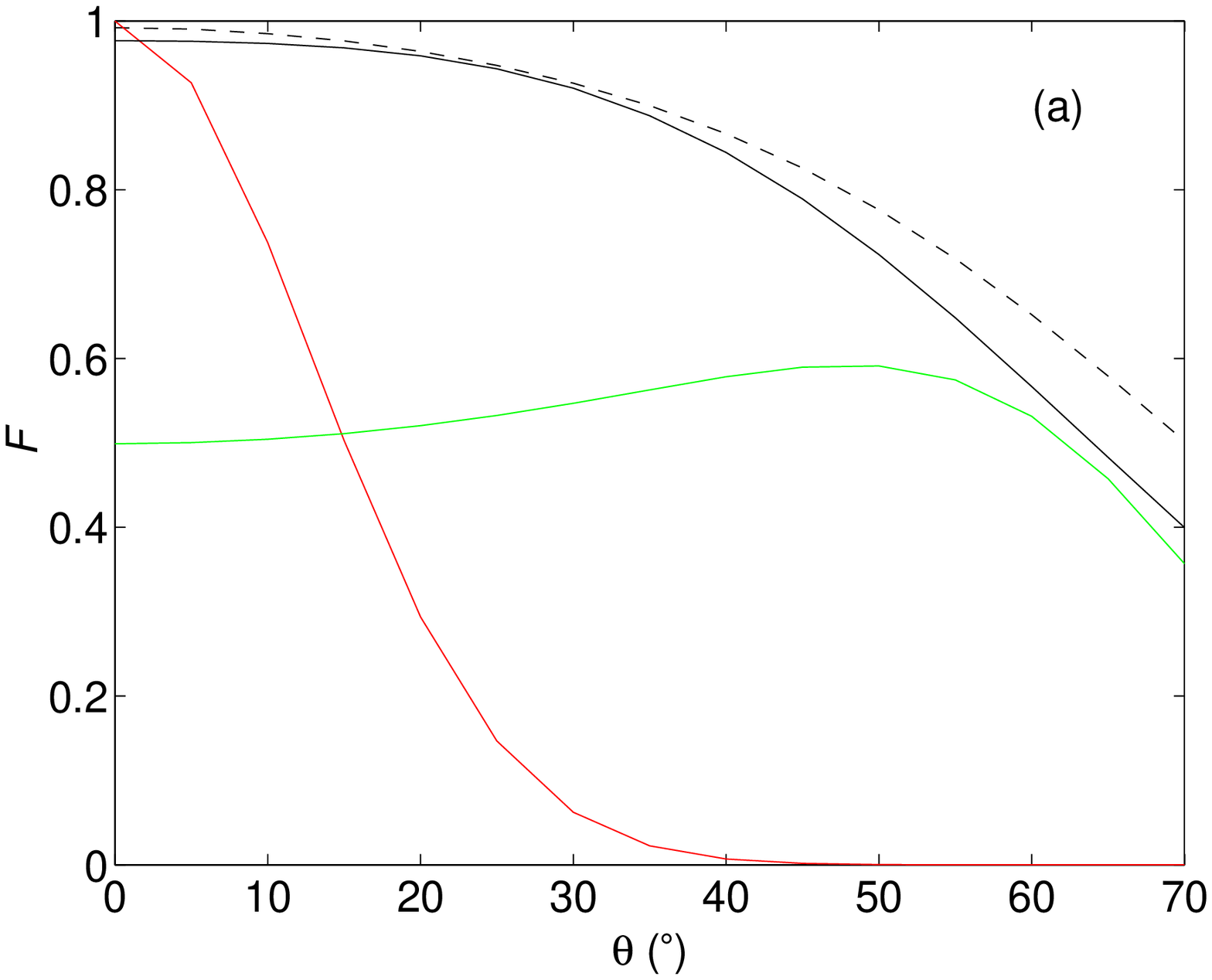}
\includegraphics[width=0.45\textwidth]{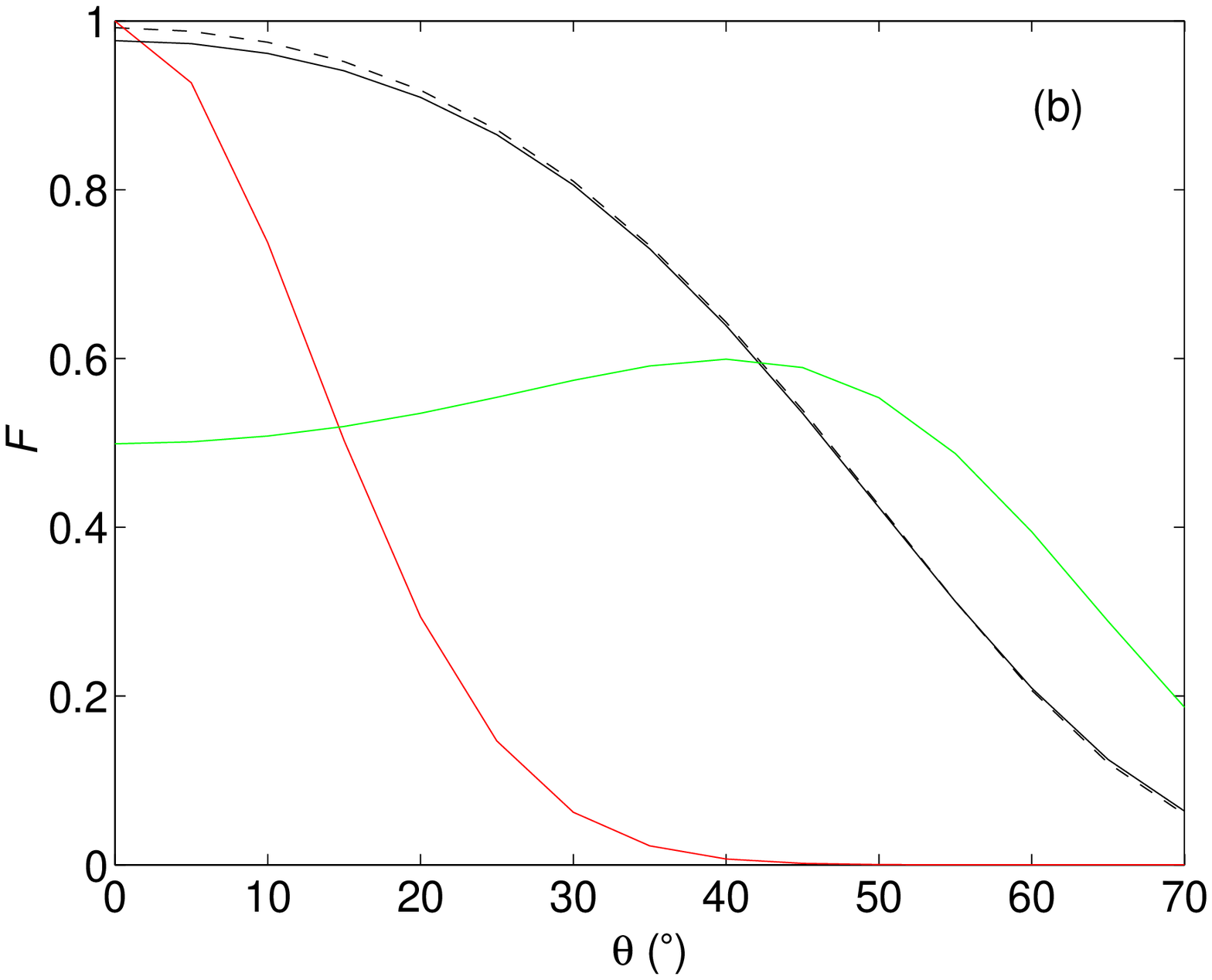}
\caption{Fidelity for teleportation of $\ket{jj}_x$ states rotated about the
$y$-axis (a) and $z$-axis (b) for $j=20$. The case for the transformation of
equation (\ref{simple}) is shown as the continuous black curve, and that using
(\ref{explicit}) is shown as the dashed curve. The green curve is for no
entanglement, and the red curve is the fidelity if the output state is always
$\ket{jj}_x$.}
\label{angy}
\end{figure}

The explicit variation of the fidelity for coherent spin states rotated by angle
$\theta$ from the $\ket{jj}_x$ state (i.e.\ $\ket{\theta,\phi=0}$ or
$\ket{\theta,\phi=\pi/2}$) is shown in figure \ref{angy}. As can be seen, the
teleportation is fairly insensitive to rotations about the $y$-axis, with high
fidelity for rotations up to $30^\circ$ or $40^\circ$. In contrast the
teleportation is more sensitive to rotations about the $z$-axis, with the
fidelity dropping off beyond about $20^\circ$. This case is for a spin of
$j=20$, but the results are similar for other spins above about 5.

In these figures the fidelity is also compared with two cases where no QT is
performed. The first alternative to QT is the fidelity in the limit $\mu \to
\infty$. In this limit the entanglement resource becomes an unentangled
$\ket{jj}_x \otimes \ket{jj}_x$ state, so we are constructing the final state
based purely on classical measurement results. In this case the fidelity is
approximately 50\%. The fidelity of teleportation does not fall to this level
until quite large rotation angles $\theta$ are reached. Note that this result is
similar to the result for CV QT \cite{Braun}, where the maximum fidelity
with no entanglement resource is 50\%. Another (trivial) alternative is where
the output state is simply $\ket{jj}_x$, independent of the input state. In this
case the fidelity is unity for zero rotation, but quickly falls to very low
levels.

\begin{figure}
\centering
\includegraphics[width=0.45\textwidth]{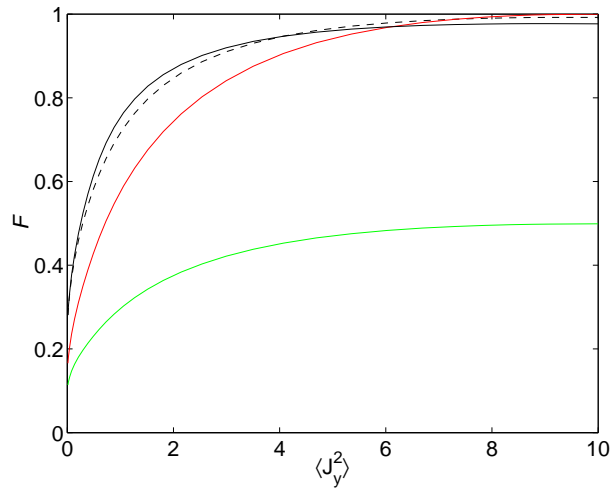}
\caption{Fidelity for teleportation of spin-squeezed states with reduced
fluctuations in $J_y$ for $j=20$. The case for the transformation of
equation (\ref{simple}) is shown as the continuous black curve, and that using
(\ref{explicit}) is shown as the dashed curve. The green curve is for no
entanglement, and the red curve is the fidelity if the output state is always
$\ket{jj}_x$.}
\label{sqzy}
\end{figure}

A more stringent way of testing the teleportation scheme is to consider input
states with non-classical features. The first example of these that we will
consider is single-mode spin-squeezed input states. More specifically, the
states that we will consider are the optimal spin-squeezed states as found using
a procedure similar to \cite{Sor01}. The results for the case of spin
squeezing in the $y$-direction are shown in figure \ref{sqzy} (the results for
squeezing in the $z$-direction are not shown as they are almost identical).
Again the fidelity for two cases where there is no teleportation has been given
for comparison. For the case where there is no entanglement, the fidelity is at
or below 50\%, and well below the fidelity for teleportation. On the other hand,
the fidelity for the case where the output state is always $\ket{jj}_x$ closely
approximates that for teleportation, except for very strong spin squeezing.

\begin{figure}
\centering
\includegraphics[width=0.45\textwidth]{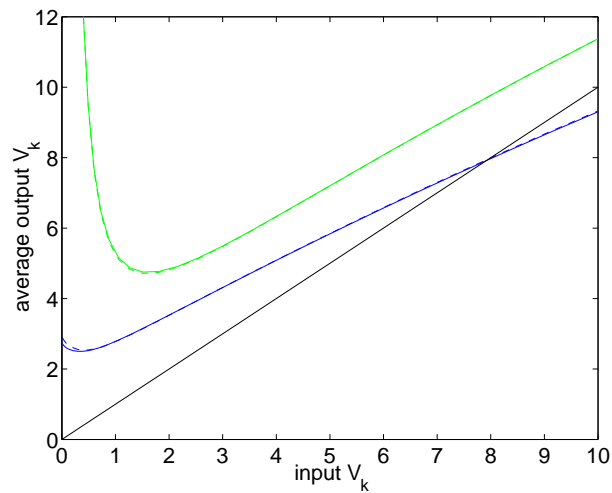}
\caption{The mean value of $V_k$ for the output state as a function of $V_k$ for
the input state for $j=20$. The results for the transformation of
equation (\ref{simple}) are shown as continuous lines, and that using
(\ref{explicit}) as dashed curves. The results for squeezing in $J_y$ and $J_z$
are shown as the green curves and blue curves respectively. The continuous black
line is that for perfect teleportation.}
\label{sqz}
\end{figure}

As the fidelity for the case where the output state is always $\ket{jj}_x$ is
about the same as that for QT, a more sensitive indication of the quality of the
teleportation is the spin squeezing in the teleported state. That is because
this indicates how much the non-classical features of the input state have been
preserved in the teleportation process. More specifically, we will consider the
quantity $V_k = {\rm Var}(J_k)$ for spin-squeezed states with reduced
fluctuations in $J_k$, $k\in \{y,z\}$. This quantity will be $j/2$ for a
coherent spin state, and less for a spin-squeezed state. For the output state
this quantity was averaged over the detection results, similar to the fidelity.
The mean value of $V_k$ for the teleported state is plotted versus the value of
$V_k$ for the input in figure \ref{sqz}.

In general the degree of squeezing of the teleported state is less than the
degree of squeezing in the original state. The preservation of squeezing is
markedly worse for low input $V_k$; this is because such a squeezed state is far
from $\ket{jj}_x$. In the case of squeezing in $J_z$, figure \ref{sqz} exhibits
an enhanced squeezing for input $V_k$ close to $j/2$. This surprising
enhancement arises because some of the squeezing that is inherent in the
two-mode squeezed resource is transferred into the output state.

\begin{figure}
\centering
\includegraphics[width=0.45\textwidth]{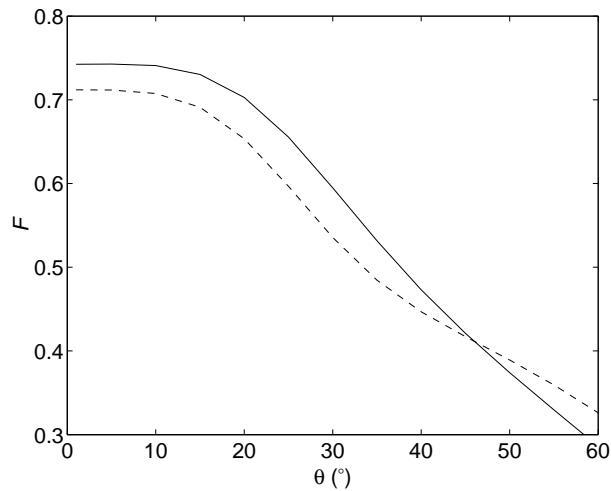}
\caption{Fidelity for teleportation of superpositions of coherent spin states
for $j=20$. The case for the transformation of equation (\ref{simple}) is shown
as the continuous curve, and that using (\ref{explicit}) is shown as the dashed
curve.}
\label{super}
\end{figure}

Another example of a state with non-classical features is a superposition of two
coherent spin states. Specifically, the states that will be considered are
\begin{equation}
\ket \psi = \exp({\rm i}\theta J_y)\ket{jj}_x -
\exp(-{\rm i}\theta J_y)\ket{jj}_x.
\end{equation}
The variation of the fidelity with $\theta$ is plotted in figure \ref{super}.
The fidelity in this case is very poor, for both types of final rotations. Note
that the fidelity is larger for smaller rotation angle. The case for zero angle
is non-physical ($\ket\psi=0$) and is not plotted. One unusual aspect of these
results is that the fidelities for the more complicated rotation
(\ref{explicit}) are below those for the simple $V(a,b)$ rotation. In contrast
the rotation of equation (\ref{explicit}) generally gives better results for
coherent spin states.

\begin{figure}
\centering
\includegraphics[width=0.45\textwidth]{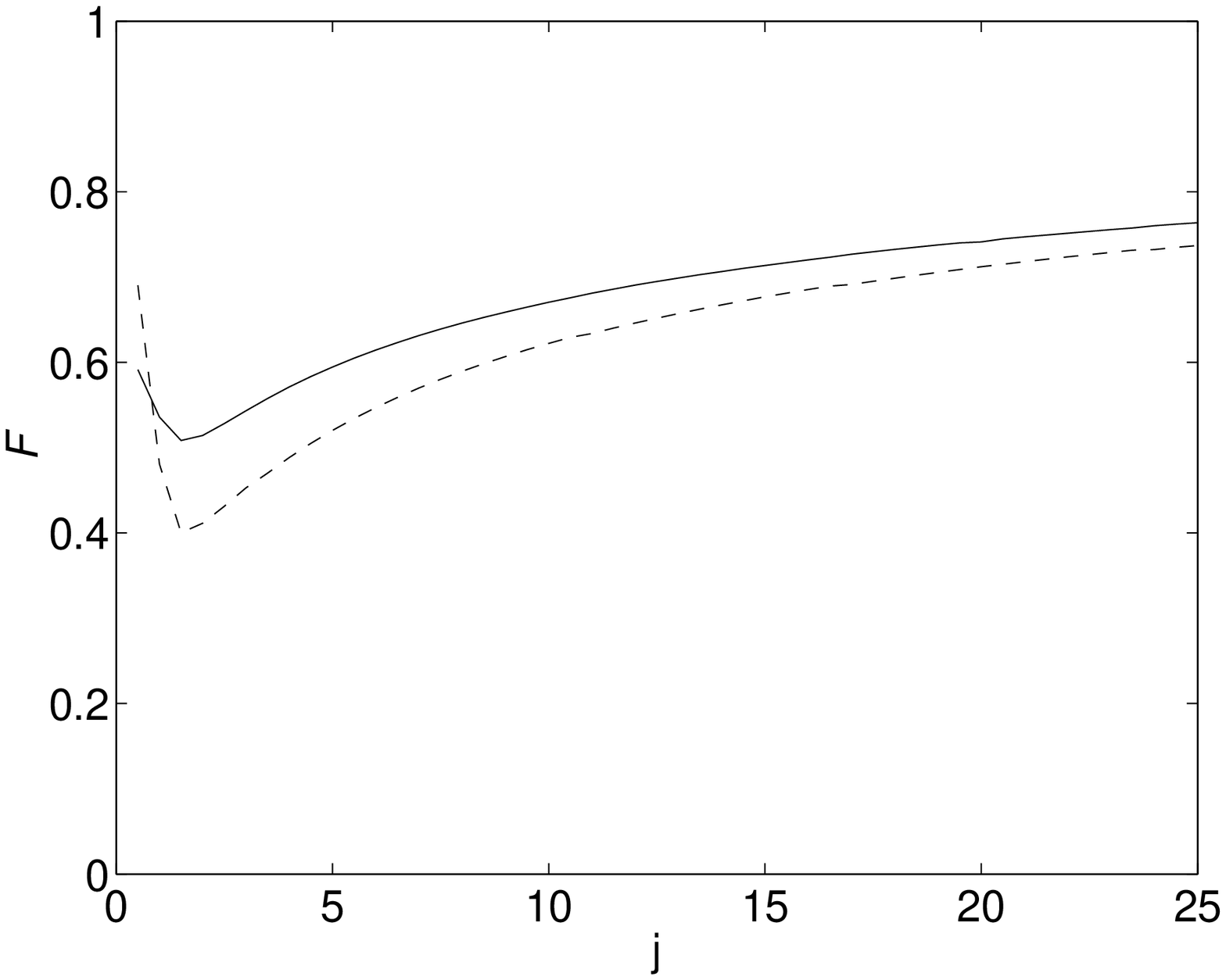}
\caption{Fidelity for teleportation of superpositions of coherent spin states
for $\theta=1^\circ$ as a function of $j$. The case for the transformation of
equation (\ref{simple}) is shown as the continuous curve, and that using
(\ref{explicit}) is shown as the dashed curve.}
\label{super2}
\end{figure}

In order for this teleportation scheme to be accurate in the limit of large
spin, we should expect the fidelity to go to 1 as the spin is increased. The
dependence of the fidelity on the spin (for a rotation angle of $1^\circ$) is
shown in figure \ref{super2}. The fidelity increases with spin for both types of
final rotation, but the fidelity is still well below 1 for the largest spin it
was feasible to perform calculations for. This low fidelity demonstrates the
fragility of a superposition of coherent states under this QT scheme.

\section{Entanglement swapping}
\label{sec:entanglement}

One way of demonstrating the quantum nature of teleportation is through ES. This
is where the state to be teleported (labelled 3), is entangled with another
state, which we will label 4. Ideal QT should teleport all properties of the
initial state, including entanglement. Therefore, the final teleported state
should be entangled with state 4. In the classical case where the final state is
reconstructed on the basis of measurement of the initial state with no
entanglement resource, there will be no entanglement between the final state and
state 4. This means that ES demonstrates that true QT has taken place.

An additional advantage of considering ES over QT is that it is independent of
the final transformation applied to the teleported state. This means that it is
possible to consider improved Bell measurements without the complication that
the fidelity is dependent on the exact final transformation applied.

Firstly we will consider the ES using the teleportation scheme considered in the
previous section. The entanglement measure that we use is entanglement of
formation:
\begin{equation}
\label{formation}
E = -{\rm Tr}(\rho_1 \log_N \rho_1).
\end{equation}
It can be shown that all measures of entanglement meeting certain basic criteria
will lie between the entanglement of distillation and the entanglement of
formation \cite{measures}. As the entanglements of distillation and formation
coincide for pure states (such as those that we are considering here), any
reasonable measure of entanglement will give the same results as equation
(\ref{formation}). Similarly to the fidelity, the entanglement was averaged over
each of the detection results, with weighting according to the probability for
obtaining those results.

\begin{figure}
\centering
\includegraphics[width=0.45\textwidth]{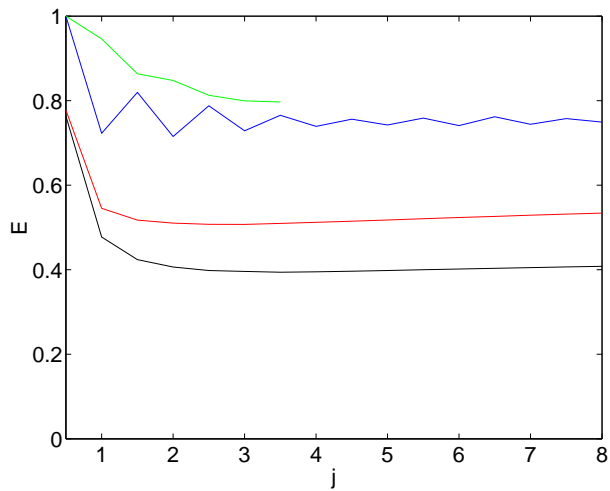}
\caption{Entanglement as a function of $j$. The ES using the
teleportation scheme of section \ref{sec:teleportation} is shown as the black
curve. The coloured curves show results for a maximally entangled entanglement
resource and three different interactions $U$: the results for the interaction
of equation (\ref{interact}) are shown as the red curve, the interaction of
equation (\ref{intalp}) as the blue curve, and the interaction of equation
(\ref{most}) as the green curve.}
\label{entang}
\end{figure}

The average entanglement using the teleportation scheme of the previous section
on maximally entangled input states is plotted in figure \ref{entang}. (We do
not show additional results for the rotations of equation (\ref{explicit}), as
the final rotations do not affect entanglement.) As can be seen, there is
significant entanglement, but the states are far less than maximally entangled.
Ideal teleportation would produce perfect ES, resulting in
the final states being maximally entangled. Nevertheless, the fact that some
ES takes place convincingly demonstrates the quantum nature
of this teleportation scheme.

When we consider ES, the values of $\mu$ found in the
previous section are far from optimum. The maximal final entanglement is
achieved for $\mu=0$, i.e.\ a maximally entangled entanglement resource. The
results for this case are also shown in figure \ref{entang}. As can be seen, the
entanglement is significantly higher than for the values of $\mu$ used in
section \ref{sec:teleportation}.

It is also possible to improve upon the ES produced by the teleportation scheme
of section (\ref{sec:teleportation}) by considering a modified Bell measurement.
The simplest alternative to consider is altering the interaction of
equation (\ref{interact}) to
\begin{equation}
\label{intalp}
U=\exp({\rm i}\alpha J_y^{(2)} J_z^{(3)}),
\end{equation}
where $\alpha$ is an arbitrary constant, rather than $1/j$. When the value of
$\alpha$ in equation (\ref{intalp}) is optimized to maximize the final
entanglement, the entanglement is as in figure \ref{entang}. The entanglement is
significantly greater than that for the interaction of equation
(\ref{interact}), but, except for the case of spin 1/2, is not unity.

In the case of spin 1/2 the ES is perfect for $\alpha=\pi$. This value of
$\alpha$ also produces perfect QT, provided that the final rotations are
${\rm e}^{{\rm i}\pi a J_z^{(3)}}{\rm e}^{-{\rm i}\pi b J_y^{(3)}}$. This gives
an alternative method to that of Bennett {\it et al} \cite{Bennett} for
achieving unit-fidelity spin 1/2 teleportation (which is equivalent to that of
Bennett {\it et al} \cite{Bennett} with the appropriate change of basis).

\begin{figure}
\centering
\includegraphics[width=0.45\textwidth]{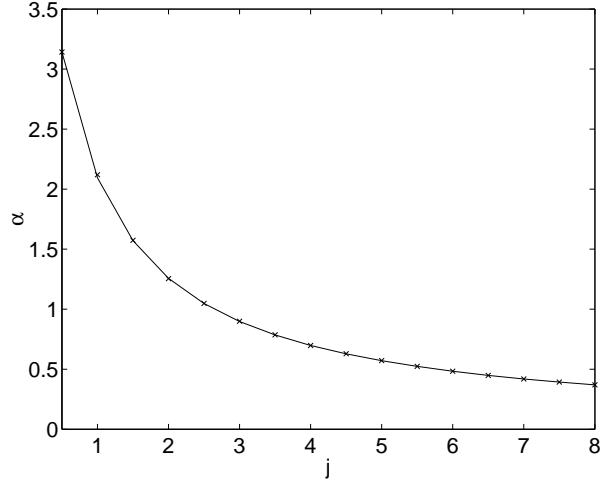}
\caption{The optimal values of $\alpha$ as a function of $j$ for the interaction
of equation (\ref{intalp}). The numerical results are shown as the crosses, and
the value of $\pi/(j+1/2)$ is shown as the continuous curve.}
\label{alpha}
\end{figure}

For $j \ge 1/2$ the optimal values of $\alpha$ are very close to $\pi/(j+1/2)$
(or $2\pi/N$), as shown in figure \ref{alpha}. In fact, for larger spins the
optimal values of $\alpha$ are virtually indistinguishable from $\pi/(j+1/2)$.
This indicates that for arbitrary spin we may improve significantly upon the
ES produced by the KP interaction, simply by changing the
value of $\alpha$ from $1/j$ to $\pi/(j+1/2)$.

Another alternative transformation, which is more general, is that given by
\begin{align}
\label{most}
&U=\exp\{{\rm i}[J_x^{(2)} (\alpha_1 J_x^{(3)}+\alpha_2 J_y^{(3)}+\alpha_3
J_z^{(3)}+\alpha_4 I^{(3)}) +J_y^{(2)} (\alpha_5 J_x^{(3)} +\alpha_6 J_y^{(3)}+
\alpha_7 J_z^{(3)}+\alpha_8 I^{(3)}) \nn \\ & +J_z^{(2)} (\alpha_9 J_x^{(3)}+
\alpha_{10} J_y^{(3)} +\alpha_{11} J_z^{(3)}+\alpha_{12} I^{(3)}) +
I^{(2)} (\alpha_{13} J_x^{(3)}+\alpha_{14} J_y^{(3)} +\alpha_{15} J_z^{(3)}+
\alpha_{16} I^{(3)})]\}.
\end{align}
Numerically solving for the optimal $\alpha_n$ gives the entanglements shown in
figure \ref{entang}. Here results are shown only up to spin $j=7/2$, as it was
not feasible to perform this numerical maximization for larger spin. As can be
seen, even allowing this far more general transformation only slightly increases
the entanglement.

Unfortunately, although these improved Bell measurements lead to improved ES,
they do not necessarily lead to improved fidelity of teleportation. The
teleportation theory based on the SU(2)$\to$HW(2) contraction does not indicate
the appropriate final rotations to perform on the teleported states for these
Bell measurements. Therefore, the final rotations considered were those given by
equation (\ref{explicit}). These rotations should approximately maximize the
fidelity for $\ket{jj}_x$, but it was found that even for these input states the
fidelity was poor.

\section{Ideal teleportation}
\label{sec:perfect}
Lastly we will show that it is possible to obtain perfect ES and QT by a minor
modification to the above scheme. As is explained in appendix \ref{append}, it
is possible to perform perfect teleportation using the transformation
${\rm e}^{\pm {\rm i}(2\pi/N) \hat n^{(2)}\hat\theta^{(3)}}$, followed by a
joint measurement of $\hat\theta^{(2)}$ and $\hat n^{(3)}$. Here the operators
$\hat\theta$ and $\hat n$ are canonically conjugate variables, and are
equivalent to the operators $J_k$ and $\theta_k$ \cite{vourdas}
(for $k\in\{x,y,z\}$) in the case of spin.

In the case of spin, we consider the phase eigenstates $\widetilde{\ket{jm}}_k$
for $k \in\{x,y,z\}$, defined by
\begin{equation}
\widetilde{\ket{jm}}_k = N^{-1/2} \sum_{n=-j}^j {\rm e}^{{\rm i} 2\pi mn/N}
\ket{jn}_k \, ,
\end{equation}
for integer spin, and by
\begin{equation}
\widetilde{\ket{jm}}_k = N^{-1/2} \sum_{n=-j}^j {\rm e}^{{\rm i} 2\pi (m+1/2)
(n+1/2)/N}\ket{jn}_k \, ,
\end{equation}
for half-odd-integer spin. These definitions are equivalent to those used in
\cite{vourdas}, though we are using different notation here. We then define
the phase operator $\theta_k$ by
\begin{equation}
\theta_k = \sum_{n=-j}^j n \widetilde{\ket{jn}}_k\widetilde{\bra{jn}} \, .
\end{equation}

This case is slightly different from that considered in appendix \ref{append}
for half-odd-integer spin; however, we still obtain identical results. To see
this, consider the Bell states given by equation (\ref{bell}), and take the
first mode to be $J_y$-eigenstates, and the second mode to be $\theta_z$-phase
eigenstates. That is, for the first mode
\begin{equation}
\ket m \equiv \ket{j,m-j-1}_y \, ,
\end{equation}
and for the second mode
\begin{equation}
\ket m \equiv \widetilde{\ket{j,m-j-1}}_z \, .
\end{equation}
Using this, the Bell states become, for integer spin,
\begin{equation}
\label{even}
\ket{p,q} = \frac 1N \sum_{n,m=-j}^j {\rm e}^{{\rm i} 2\pi mp/N}
{\rm e}^{{\rm i} 2\pi (m+q)n/N} \ket{jm}_y \ket{jn}_z \, ,
\end{equation}
and, for half-odd-integer spin
\begin{equation}
\label{odd}
\ket{p,q} = \frac 1N \!\!\! \sum_{n,m=-j}^j \!\!\!\! {\rm e}^{{\rm i} 2\pi mp/N}
{\rm e}^{{\rm i} 2\pi (m+q+1/2)(n+1/2)/N} \ket{jm}_y \ket{jn}_z .
\end{equation}
In either case the Bell states simplify to
\begin{equation}
\ket{p,q} = {\rm e}^{{\rm i} (2\pi/N) J_y^{(2)}J_z^{(3)}} \widetilde{\ket{jp}}_y
\widetilde{\ket{jq}}_z \, .
\end{equation}
This is the result that we obtain using the derivation in appendix \ref{append},
with $J_y^{(2)} \equiv \hat n^{(2)}$ and $J_z^{(3)} \equiv \hat \theta^{(3)}$.
Similarly as in appendix \ref{append}, we can also obtain teleportation using
the negative sign in the exponential. We can derive this result from the Bell
states in the form
\begin{equation}
\ket{p,q} = N^{-1/2} \sum_{m=1}^N {\rm e}^{{\rm i} 2\pi mp/N} \ket m
\ket{q-m\!\!\!\!\mod N}\, ,
\end{equation}
rather than equation (\ref{bell}).

These results mean that it is possible to perform these ideal Bell measurements
by using the transformation ${\rm e}^{\pm {\rm i} (2\pi/N) J_y^{(2)}J_z^{(3)}}$
followed by measurements of $\theta_y^{(2)}$ and $\theta_z^{(3)}$. This is very
similar to the result that was found in the previous section, that effective
ES may be achieved using the transformation ${\rm e}^{{\rm i}(2\pi/N) J_y^{(2)}
J_z^{(3)}}$, followed by spin measurements.

Thus we find that it is possible to perform perfect teleportation (and therefore
ES) that is equivalent to that considered by Bennett {\it et al} \cite{Bennett},
by using a slightly
different unitary transformation than that considered by KP, and replacing the
spin measurements with phase measurements. The modification to the unitary
transformation is trivial, and is equivalent to applying the interaction for a
longer time. Unfortunately the measurement of phase is non-trivial, and it is
not, in general, possible to perform these measurements by simple rotations and
spin measurements. The only exception to this is the spin-$1/2$ case, as the
phase eigenstates are also spin eigenstates \cite{qudit}. This is why it is
possible to perform perfect QT and ES in the spin-$1/2$ case.

Nevertheless, one of the main applications of this QT is the
transport of states between quantum computers based on qudits. One of the
requirements for construction of such a quantum computer is that it is possible
to perform arbitrary unitary transformations on a single mode \cite{qudit}. In
that case it is possible to perform phase measurements, simply by performing a
unitary transformation between the basis of phase states and the basis of spin
states.

It is interesting that the unitary transformation required for perfect
teleportation here is distinct from that for the approximate KP teleportation,
and does not approach it in the limit of large $j$. As is shown in
appendix~\ref{limitN}, the perfect teleportation considered here is equivalent
to CV teleportation in the limit of large $j$. Similarly the
teleportation scheme of section \ref{sec:teleportation} is equivalent to
CV teleportation in the limit of large $j$ for states close to
the maximally weighted $J_x$-eigenstate. Nevertheless, these limits are
fundamentally different, and it does not appear to be possible to perform
teleportation that is perfect for arbitrary input states, but is equivalent to
KP teleportation for states near $\ket{jj}_x$.

\section{Conclusions}
\label{sec:conclusions}
Quantum teleportation has not yet been experimentally demonstrated for finite
spin above $1/2$. Here we have shown that, using the Ising interaction, spin
rotations and spin measurements, it is possible to perform QT of states close
to $\ket{jj}_x$. For coherent spin states that are not rotated by more than
about $20^\circ$ from $\ket{jj}_x$ the fidelity of teleportation is higher than
90\%. The fact that the fidelity is not exactly unity is not a great drawback
in achieving an experimental realization, as it is still much higher than what
is currently achievable experimentally. In addition it is more than sufficient
to demonstrate that true QT is taking place, in that the
fidelity is higher than what would be possible without entanglement.

This teleportation scheme also teleports the non-classical features of input
states. For input spin-squeezed states, the teleported state also exhibits
spin squeezing. In addition, a superposition of two coherent spin states may be
teleported, and if the input state is entangled with another state, some of the
entanglement will be teleported. The teleportation of these non-classical
features is generally poorer, however. The fidelity for teleportation of a
superposition of coherent spin states is poor, except for very large spin and
small separation between the two coherent states in the superposition. Also,
only about 40\% of the entanglement is teleported.

There are improvements that can be made on the ES given by this teleportation
scheme. One improvement that can be made is to simply use an optimally entangled
entanglement resource. Further improved ES can be obtained by modifying the
unitary transformation used for KP Bell measurements from ${\rm e}^{{\rm i}(1/j)
J_y^{(2)}J_z^{(3)}}$ to ${\rm e}^{{\rm i} (2\pi/N) J_y^{(2)}J_z^{(3)}}$. In the
case of spin $j=1/2$, this interaction also allows perfect QT.

If the Bell measurements are further modified by considering phase measurements
rather than spin measurements, then it is possible to achieve perfect QT and
ES. These Bell measurements are equivalent to those considered by Bennett
{\it et al} \cite{Bennett}
under the appropriate change of basis. Performing phase measurements is
non-trivial, and cannot be done using rotations and spin measurements.
Nevertheless, if it is possible to perform the more general single-mode unitary
transformations required for a quantum computer \cite{qudit}, it should be
possible to perform QT in this way.

\acknowledgments
We gratefully acknowledge valuable discussions with S D Bartlett. This
research was supported by the Australian Research Council.

\appendix
\section{Fidelity}
\label{apfid}
In this paper we consider three different types of fidelity. The usual
definition of fidelity is \cite{Sch96}
\begin{equation}
{\cal F} = \bra \psi \rho \ket \psi ,
\end{equation}
where $\rho$ is Bob's output density operator which can, in general, be mixed.
Here we consider Bob's output state $\ket{\zeta_{a,b}}$ to be a pure state that
is dependent on the Bell measurement results $(a,b)$ for QT. Using the above
definition we obtain a fidelity that is dependent on the input state
$\ket\psi$ and the measurement results $(a,b)$:
\begin{equation}
{\cal F}(\ket \psi, a, b) \equiv |\braket{\psi}{\zeta_{a,b}}|^2 .
\end{equation}
We generally do not use this expression as we wish to know the unconditional
fidelity, averaged over the measurement results:
\begin{equation}
\label{eq:fidelity}
{\cal F}(\ket \psi) \equiv \sum_{a,b} P(a,b|\psi)
|\braket{\psi}{\zeta_{a,b}}|^2 \, ,
\end{equation}
where $P(a,b|\psi)$ is the probability of Alice obtaining the measurement
results $(a,b)$. This expression is equivalent to that considered for schemes
without entanglement in \cite{Hender}.

The fidelity given by equation (\ref{eq:fidelity}) is applicable only for a
specific input state, for which the teleportation is trivial. In general we wish
to consider teleportation of a range of states. In this case it is more
appropriate to determine the fidelity averaged over this range of states, using
\begin{equation}
\label{fidav}
{\cal F}_{\rm av} = \int W(\ket\psi) \left[ \sum_{a,b} P(a,b|\psi) |\braket
{\psi}{\zeta_{a,b}}|^2 \right] {\rm d}\psi \, ,
\end{equation}
where $W(\ket\psi)$ is the weighting function over the set of states $\{\ket
\psi\}$. We use the differential d$\psi$ to indicate an integral over the state
coefficients with the additional constraint that the state is normalized, i.e.
\begin{equation}
{\rm d}\psi \equiv \delta(|\braket \psi \psi |-1){\rm d}^{2N} (\braket m \psi),
\end{equation}
where $\ket m$ is a basis state of ${\cal H}_N$ as used in equation
(\ref{bell}). In this study, we consider a weighted average over coherent spin
states; that is
\begin{equation}
{\cal F}_{\rm av} = \int W(\ket{\theta,\phi})\left[ \sum_{a,b} P(a,b|\theta,
\phi) |\braket{\theta,\phi}{\zeta_{a,b}}|^2 \right] {\rm d}\Omega \, ,
\end{equation}
where d$\Omega$ is a unit of solid angle and $\ket{\theta,\phi}$ is a coherent
spin state, as in equation (\ref{defcoh}).

\section{Bell states}
\label{append}
We can consider Bell states slightly more general than those of equation
(\ref{bell}), of the form 
\begin{equation}
\ket{p,q,s_1,s_2,s_3} = N^{-1/2} \sum_{m=1}^N {\rm e}^{{\rm i} s_1 2\pi mp/N}
\ket m \ket{s_2 m + s_3 q},
\end{equation}
where the variables $s_k$, $k\in\{ 1,2,3\}$, take the values $\pm 1$, and the
modulo $N$ has been omitted for brevity. Each of these states is maximally
entangled, and states for differing $p$ or $q$ are orthogonal. Now we define the
conjugate states
\begin{equation}
\label{fourier}
\widetilde{\ket m}=N^{-1/2} \sum_{n=1}^N {\rm e}^{{\rm i} s_k 2\pi mn/N}\ket n .
\end{equation}
For additional generality we have included the signs $s_k$. We will use the
subscripts $k=4$ and 5 for modes 2 and 3, respectively. The states $\ket m$ and
$\widetilde{\ket n}$ satisfy the conjugacy relation
\begin{equation}
\langle m \widetilde{\ket n} = N^{-1/2} {\rm e}^{{\rm i} s_k 2\pi mn/N} ,
\end{equation}
analogous to the conjugacy relation between position and momentum eigenstates,
$\ket x$ and $\ket p$, respectively:
\begin{equation}
\braket xp = (2\pi)^{-1/2} {\rm e}^{{\rm i}xp} ,
\end{equation}
where we have taken $\hbar=1$. By using the inverse relation to (\ref{fourier})
on mode 3,
\begin{equation}
\ket n=N^{-1/2} \sum_{n=1}^N {\rm e}^{-{\rm i} s_5 2\pi mn/N}\widetilde{\ket m},
\end{equation}
we obtain
\begin{equation}
\ket{p,q,s_1,s_2,s_3}=\frac 1N \sum_{n,m=1}^N {\rm e}^{{\rm i} s_1 2\pi mp/N}
{\rm e}^{-{\rm i} s_5 2\pi n(s_2 m + s_3 q)/N}\ket m \widetilde{\ket n}.
\end{equation}
This can be expressed as
\begin{equation}
\ket{p,q,s_1,s_2,s_3} = {\rm e}^{-{\rm i} s_5 s_2 2\pi \hat n^{(2)}
\hat\theta^{(3)}/N}\widetilde{\ket{s_4 s_1 p}}\ket{s_3 q} ,
\end{equation}
where the operators $\hat n$ and $\hat \theta$ denote the operators
corresponding to the states $\ket n$ and $\widetilde{\ket n}$, respectively, the
superscripts $(2)$ and $(3)$ indicate operators on the first and second modes,
respectively, and the modulo $N$ has been omitted from the states for brevity.

This demonstrates that perfect teleportation can be achieved by performing the
transformation ${\rm e}^{\pm (2\pi {\rm i}/N) \hat n^{(2)}\hat\theta^{(3)}}$,
followed
by a joint measurement of $\hat\theta^{(2)}$ and $\hat n^{(3)}$. Note that the
sign in this transformation is arbitrary, and does not depend on the signs used
in transforming between the conjugate states, as there is the additional sign
$s_2$. In more general terms, teleportation is achieved when the operators in
the interaction are conjugate to those that are measured.

\section{Large-$N$ limit}
\label{limitN}
Considering the Bell states given by equation (\ref{bell}), it is clear that the
result $q$ is equivalent to the result of a measurement of $\hat n^{(3)}-\hat
n^{(2)}$ modulo $N$. It is possible to re-express the Bell states (\ref{bell})
as
\begin{equation}
\ket{p,q} = N^{-1/2} \sum_{m=1}^N {\rm e}^{-{\rm i} 2\pi nq/N}
\widetilde{\ket{p-n}}\widetilde{\ket n} \, ,
\end{equation}
where the modulo $N$ has again been omitted for brevity. Here we are using the
conjugate states as in the previous section with $s_k=1$, $k\in\{4,5\}$. From
this form of the Bell states, it is clear that $p$ is equivalent to a
measurement of $\hat \theta^{(2)}+\hat \theta^{(3)}$ modulo $N$.

Thus we find that the Bell measurements are equivalent to a joint measurement of
$\hat n^{(3)}-\hat n^{(2)}$ and $\hat \theta^{(2)}+\hat \theta^{(3)}$ modulo
$N$. In addition, the entangled state is equivalent to an eigenstate of $\hat
n^{(2)}-\hat n^{(1)}$ and $\hat \theta^{(1)}+\hat \theta^{(2)}$ modulo $N$. This
is very similar to the case of teleportation of continuous variables
\cite{continuous}.

Let us make the substitutions
\begin{equation}
\hat x_N= \sqrt{\frac{2\pi}N}\, \hat n \, , ~~~ \hat p_N=\sqrt{\frac{2\pi}N}\,
\hat \theta \, .
\end{equation}
The eigenvectors for these variables satisfy
\begin{equation}
\braket{x_N}{p_N} = N^{-1/2} {\rm e}^{{\rm i}x_N p_N},
\end{equation}
which is equivalent to that for position and momentum, apart from a multiplying
factor. In the limit $N \to \infty$, the ranges of $x_N$ and $p_N$ go to
infinity, while the spacing between the eigenvalues goes to zero, so these
variables are equivalent to continuous position and momentum. In this limit, it
is clear that the teleportation is equivalent to the teleportation considered in
\cite{continuous}.


\begin{thebibliography}{}
\bibitem{randsta} The initial random state is assumed to have real coefficients
$\Phi_{mn}$ that are selected from independent normal distributions, centred at
zero for each $(m,n)$, with the constraint that the state $\ket \Phi$ is
normalized.
\end{thebibliography}
\end{document}